\documentstyle[12pt,epsfig]{elsart}

\date{}
\def\Journal#1#2#3#4{{#1} {\bf #2}, #3 (#4)}

\def\NIMA{{\em Nucl. Instrum. Methods} A}

\def\NDS{{\em Nucl. Data Sheets} } 
\def\PLB{{\em Phys. Lett.} B}
\def\PRP{{\em Phys. Rep.}}
 
\def\PRL{\em Phys. Rev. Lett.} 
 
\def\PRC{{\em Phys. Rev.} C}

\def\PNPP{\em Prog. Nucl. Part. Phys.}  
 
\def\EPA{{\em Europ. Phys. J.} A} 
 
\def\APJ{\em ApJ}

\def\ra{\rightarrow} 
\def\be{\begin{equation}} 
\def\ee{\end{equation}}
\def\bea{\begin{eqnarray}} 
\def\eea{\end{eqnarray}}

\newcommand{\bmbm}{\mbox{$\beta^-\beta^-$} }
\newcommand{\bpbp}{\mbox{$\beta^+\beta^+$} }

\newcommand{\enu}{\mbox{$E_{\nu}$} }

\newcommand{\nel}{\mbox{$\nu_e$} }

\newcommand{\neu}{neutrino }
\newcommand{\bdec}{$\beta$-decay} 
\newcommand{\neus}{neutrinos }

\newcommand{\zns}{\mbox{$^{70}Zn$ }}
\newcommand{\gas}{\mbox{$^{70}$Ga }}
\newcommand{\gaes}{\mbox{$^{71}$Ga }}
\newcommand{\ges}{\mbox{$^{70}$Ge }}
\newcommand{\bes}{\mbox{$^7$Be }} 
\newcommand{\clsd}{\mbox{$^{37}$Cl }}
\newcommand{\yhss}{\mbox{$^{176}$Yb }}   
\newcommand{\moeh}{\mbox{$^{100}$Mo }}

\newcommand{\cd}{\mbox{$^{116}Cd$ }} 
\newcommand{\inhs}{\mbox{$^{116}In$ }} 
\newcommand{\cdhdz}{\mbox{$^{113}Cd$ }}
\newcommand{\inhdz}{\mbox{$^{113}In$ }}
\newcommand{\inhf}{\mbox{$^{115}In$ }}
\newcommand{\cdhs}{\mbox{$^{106}Cd$ }} 

\newcommand{\cdhvz}{\mbox{$^{114}Cd$ }} 

\newcommand{\gdhs}{\mbox{$^{160}Gd$} }
 
\newcommand{\teha}{\mbox{$^{128}Te$ }} 
 
\newcommand{\tehfz}{\mbox{$^{125}Te$ }} 
\newcommand{\ihfz}{\mbox{$^{125}I$ }} 
\newcommand{\tehd}{\mbox{$^{130}Te$ }}
\newcommand{\xehd}{\mbox{$^{130}Xe$ }}
\newcommand{\ihd}{\mbox{$^{130}I$ }}

\newcommand{\bnel}{\mbox{$\bar{\nu}_e$} }

\begin{document}
\begin{frontmatter} 
\title{Spectroscopy of low energy solar neutrinos using CdTe detectors} 
\author{K. Zuber} 
\address{Oxford University, Dept. of Physics, Denys Wilkinson Building,\\ 
Keble Road, Oxford OX1 3RH, England} 
\begin{abstract} 
The usage of a large amount of CdTe(CdZnTe) semiconductor detectors for
solar neutrino
spectroscopy in the low energy region is investigated.
Several different coincidence signals
can be formed on five different isotopes to measure the \bes neutrino line
at 862 keV in real-time. The most promising one is the usage of \cd resulting
in 89 SNU. 
The presence of \tehfz permits even the real-time detection of
pp-neutrinos. 
A possible antineutrino flux above 713 keV might be detected by capture on
\cdhs . 
\end{abstract} 
\end{frontmatter} 

\section{Introduction} 
Over the last years striking evidence arose for a non-vanishing neutrino
rest mass (for reviews see \cite{zub98,samoil}). They all come from
neutrino oscillations experiments.
Among them is the long standing evidence of a solar neutrino deficit also being
of fundamental importance for stellar astrophysics. The deficit is seen
in radiochemical detectors, namely GALLEX/GNO \cite{gallex,kirsten} and SAGE \cite{gavrin}
using \gaes , still the only pp-neutrino detectors available, and the
Homestake experiment using \clsd \cite{ray}. A reduced $^8$B \nel flux is measured by
two water Cerenkov detectors, namely Super-Kamiokande \cite{smy} and SNO \cite{snocc}. A difference in
measured
fluxes among the 
latter resulted in evidence for an active
neutrino flavour coming from the sun besides \nel.  
This is due to the fact that Super-Kamiokande is using
neutrino-electron scattering and SNO inverse
\bdec{} for detection. Recent SNO results of neutral current reactions on deuterium dramatically confirm the existence
of further active neutrinos, being the dominant solar neutrino flux 
\cite{snonc}.
The solution of the solar neutrino deficit has to come from particle physics,
the scenario discussed most often is neutrino oscillations. 
Taking all experimental results together the largest effects in the solar neutrino spectrum 
implied by the various oscillations solutions show up in the region below
1 MeV. Furthermore this region corresponds to 99 \%
 of the solar neutrino flux and is still very
important for understanding stellar energy generation \cite{bah02}.
An interesting idea to measure such low energy neutrinos in real time is suggested
by \cite{rajulens} using coincidence techniques for neutrino capture on nuclei. 
It is already finding its practical application in the LENS \cite{lens}, SIREN \cite{siren} and MOON projects \cite{moon}.
aiming to measure pp-neutrinos in real time.
The technique relies on using either a large amount of double beta isotopes (\yhss in case of LENS, \gdhs in case
of SIREN and \moeh
in case of MOON)
or highly forbidden beta decay emitters like \inhf (4-fold forbidden, currently under study in LENS as an alternative
to \yhss ), as target material \cite{rajulensin}. Clearly an interesting spin off is the investigation of double beta
decay \cite{zub00}.\\ 
In this paper the possibility to apply the same technique 
for CdTe(CdZnTe) semiconductor detectors and their feasibility for solar
neutrino
detection is explored. CdTe semiconductor detectors
have already a wide field of application in $\gamma$-ray astronomy and medical physics.
The study performed here is motivated by the COBRA project \cite{cobra}, planning to use large amounts of CdTe-detectors for double beta
decay searches. The usage of large amounts of semiconductors for solar neutrino detection was also considered in the past
for Ge-detectors \cite{laura} and GaAs \cite{bowles} relying largely on 
the detection of electrons from neutrino-electron scattering. 
In the case discussed here, we focus on the detection of \bes and
pp-neutrinos
only.
Measurements of higher energetic neutrino flux components are also
possible but will not be discussed. Also a possible
real-time detection of pp-neutrinos via neutrino-electron scattering as
well as contributions from the CNO cycle are not considered.

\section{Solar neutrino detection via coincidence measurements}

The detection principle for solar neutrinos using coincidences relies on the following two reactions
\bea
\label{eq:gs}
\nel + (A,Z) \ra (A,Z+1)_{g.s.} + e^- \ra (A,Z+2) + e^- + \bnel \\
\label{eq:es}
\nel + (A,Z) \ra (A,Z+1)^\ast + e^-  \ra (A,Z+1)_{g.s.} + \gamma 
\eea
Therefore either coincidence between two electrons for the ground state transitions or the
coincidence of an electron with the corresponding de-excitation photon(s) is
required.
The first one is followed by MOON, while the second one is used for LENS.
The produced electrons as the first part of the coincidence
have energies of
\begin{equation}
E_e = E_{\nu} - (E_f - E_i)
\end{equation}
with \enu as neutrino energy, $E_f$ and $E_i$ as the energy of the final 
and initial nuclear state involved in the transition.
In case of \bes \neus the electrons will be monoenergetic.

\subsection{Solar \neu tags using double beta isotopes}
\begin{figure}
\begin{center}
\epsfig{file=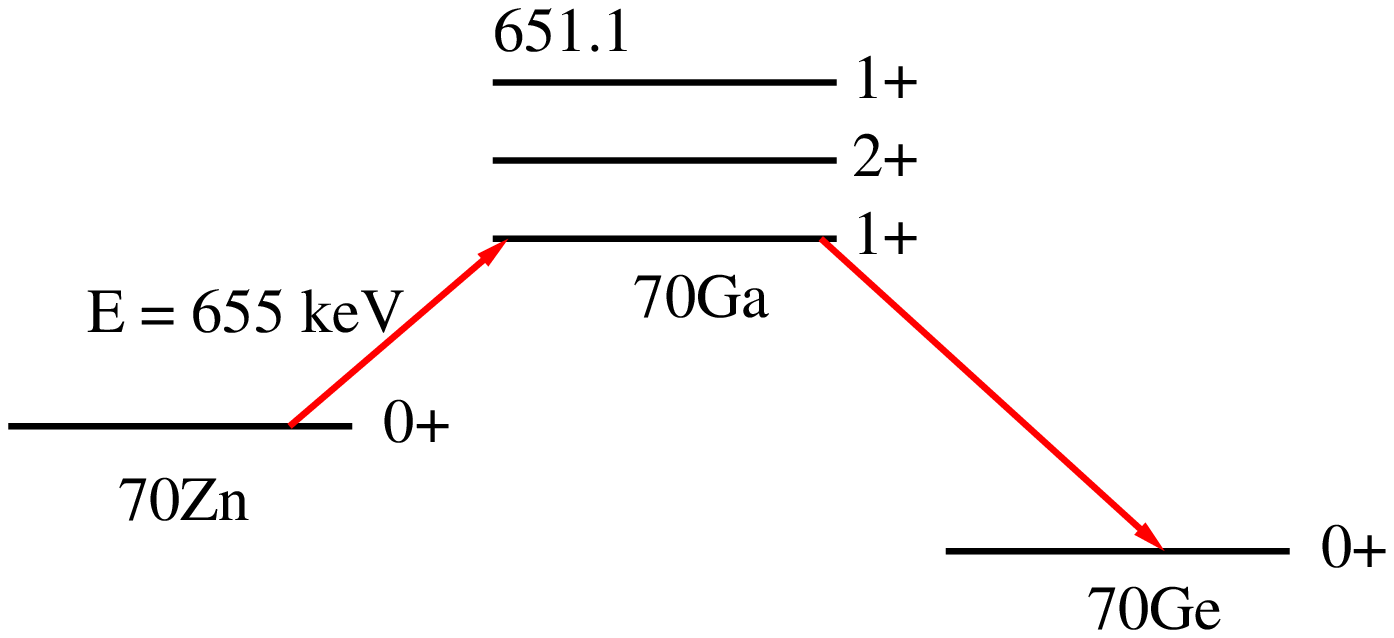,width=5cm,height=3cm}
\epsfig{file=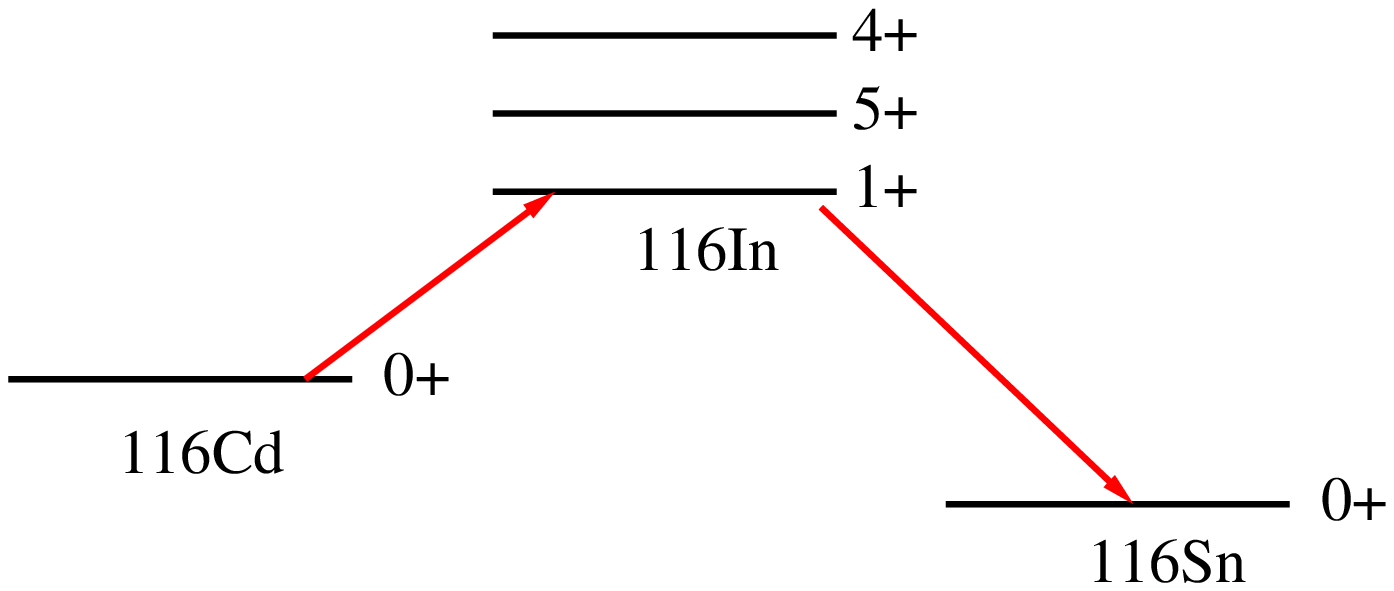,width=5cm,height=3cm}
\epsfig{file=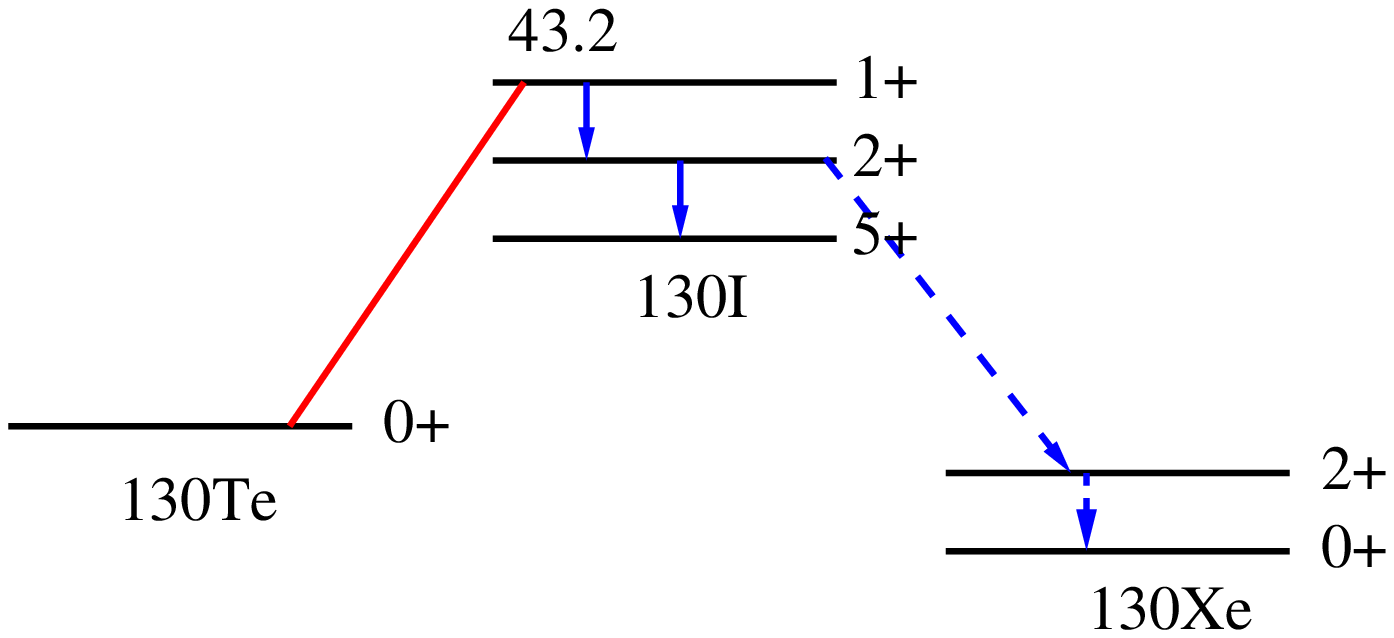,width=5cm,height=3cm}

\end{center}
\caption{\label{pic:bbtags} Possible \bes neutrino tags using \zns (left), \cd (middle) and \tehd
(right). 
The most promising one is \cd, having a threshold of 464 keV for solar neutrinos and a short
coincidence time between the two electrons, because the half-life of \inhs is only 14.1 s. The
metastable 2$^+$ state made decay via IT or \bdec{} to the first excited 2$^+$ state of \xehd.}
\label{fig6}
\end{figure}

Consider detection with the help of double beta isotopes first.
There are 4 (5) \bmbm emitters in CdTe (CdZnTe) \cite{cobra}.
Three of them have sensitivity to the \bes line of 862 keV, namely \zns , \cdhs and \tehd .
The corresponding coincidence tags are shown in Fig.~\ref{pic:bbtags}. 
\zns and \cd are in the form of ground state
transitions (Eq.~\ref{eq:gs}), while for \tehd an e-$\gamma$ coincidence
(Eq.~\ref{eq:es}) is required. 
In case of \zns it is the ground state transition to \gas with a threshold of \enu = 655 keV. \gas will decay via beta
decay to the ground state of \ges in 98.9 \% of all cases
and a half-life of 21.14 min.
However because \zns is only present in 
small amounts (Zn typically replaces 10 \% 
of Cd in CdZnTe detectors) in the detector, the expected rate is rather small.
In case of \tehd it is the transition to the first excited 1$^+$ state in \ihd, lying 43.25 keV above ground state. 
This corresponds to a neutrino energy threshold of \enu = 494 keV. The state de-excites under the emission of a 3.3 keV
X-ray. With a half-life of 8.8 mins in 86 \% of the cases an IT will happen resulting in a 39.95 keV photon or in 
14 \% a \bdec{} to \xehd, dominantly inot the first excited 2$^+$ state. This is connected with the emission of a 536
keV photon. 
The analogous \bdec{} of the 5$^+$ ground state of \ihd with a half-life of 12.36 h is probably inadequate to use in the coincidence.
Two more 1$^+$ states exist at 254.8 keV and 349.6 keV above ground state which can be populated by \bes, corresponding to a neutrino energy
threshold of \enu = 706 keV and 801 keV respectively. Furthermore there exist several low-lying states, whose quantum numbers have
not been determined yet and it might be worthwile to do so.  
Probably the most appropriate isotope for the search is \cd.
The coincidence signal will be the two electrons of the neutrino capture to the ground state of \inhs 
with an energy threshold of \enu = 464 keV together with the \bdec{} (practically 100 \%) of
\inhs with a half life of 14.1 s.
The Q-value of the latter is 3.275 MeV.
Thus a good coincidence signal can be formed among the
two electrons. Such double electron tags can also be performed for
detecting the pep-line at 1.445 MeV by using \cdhvz. Here a low energy electron of
about
5 keV is in coincidence with a \bdec{} electron of $^{114}$In, having a
half-life of 72s and a Q-value of 1.98 MeV.
 
\subsection{The case of \cdhdz}
\begin{figure}
\begin{center}
\epsfig{file=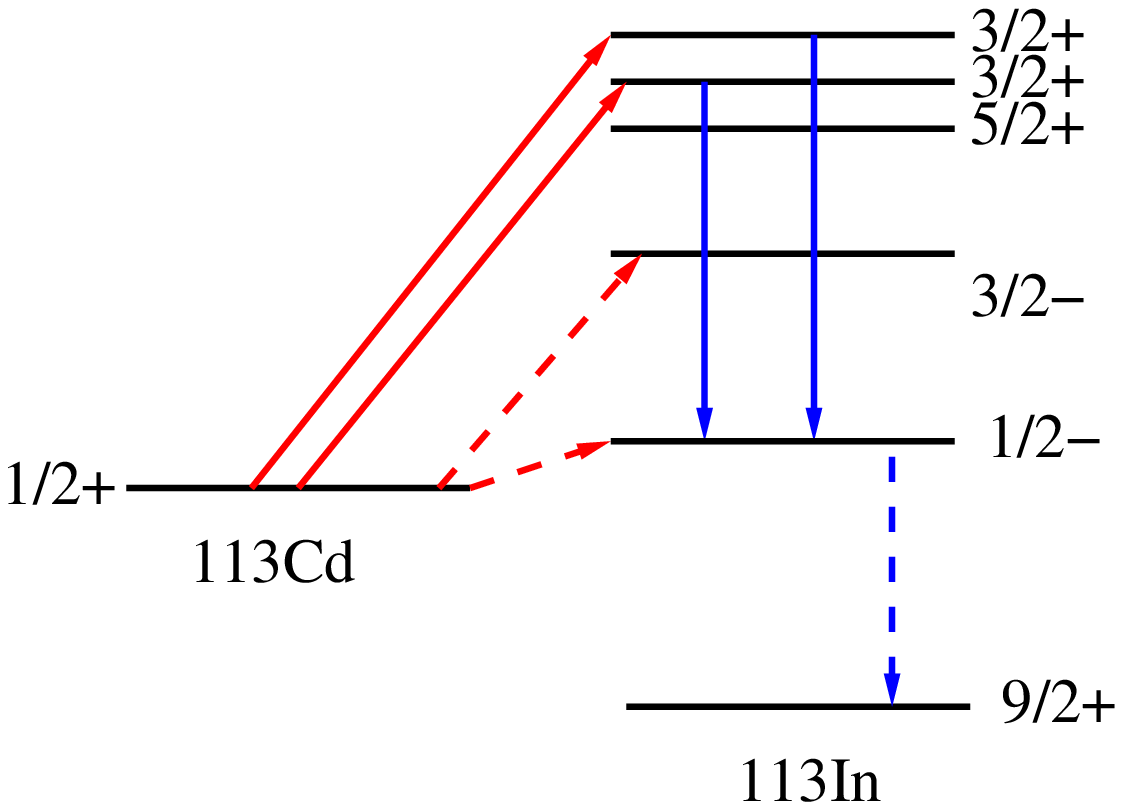,width=5cm,height=3cm}
\epsfig{file=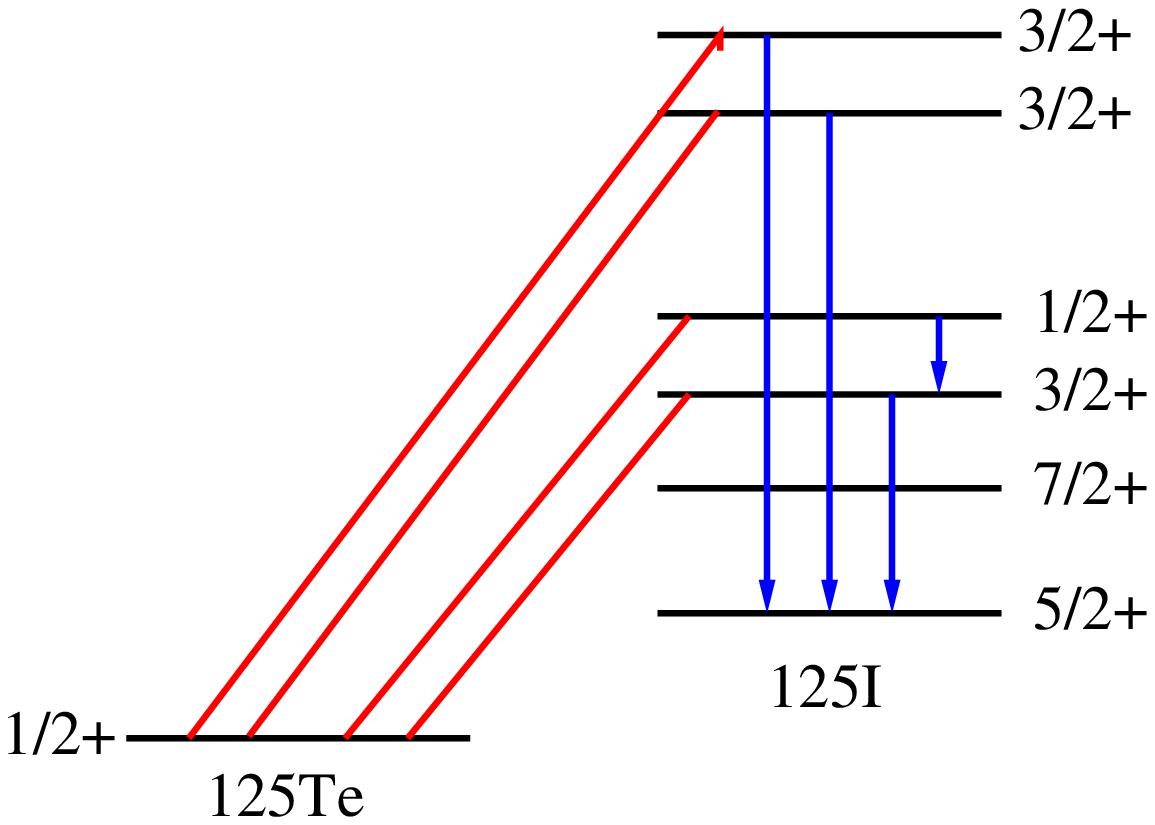,width=5cm,height=3cm}
\end{center}
\caption{\label{pic:otags} Left: The various solar neutrino captures on \cdhdz. Two allowed
transitions exist for \bes \neus 
(solid lines). 
Two forbidden transition (dashed lines) would allow the detection of pp-neutrinos with a threshold of
\enu = 70 keV only, however the
involved GT matrix elements
are orders of magnitude lower than for allowed transitions. Right: \tehfz allows the detection of pp and \bes neutrinos in
real time. The two low lying 1/2$^+$ and 3/2$^+$ state can be populated by pp-neutrinos. The threshold is 330 keV.}
\label{fig6}
\end{figure}

One might also consider the case of \cdhdz as target for a neutrino capture into \inhdz. \cdhdz is
one out of only three known 4-fold forbidden beta decay isotopes, besides
\inhf and $^{50}$V. Indeed, since quite some time the idea to build an
indium
solar neutrino detector exists \cite{raju1}, a possible realisation is
considered now within the 
LENS experiment. It is therefore 
natural to ask, whether \cdhdz is also interesting.
The possible solar neutrino tag is shown in Fig.~\ref{pic:otags}. 
As can be seen, also with this isotope \bes spectroscopy of the sun is possible. 
There are two excited 3/2$^+$ states  at 1029.6 and 1063.9 keV above ground state, resulting
in neutrino thresholds of \enu = 709 and 743 keV respectively. The
electron is most of the time accompanied by
a 672 or 638 keV photon. The IT of the 1/2$^-$ state to the 9/2$^+$ ground
state of \inhdz is associated with an additional 391.7 keV photon and a
half-life of 1.65 h. Two forbidden transitions exist which in principle
would allow a real time-detection of pp-neutrinos with a threshold of
\enu = 70 keV and 330 keV, however the expected smallness of the involved
nuclear matrix elements will result in a rather small rate.\\
Possible advantages with respect to \inhf are that the beta half-life of
\cdhdz is more than one order
of magnitude higher than for \inhf, resulting in correspondingly less background from that process
(typical background from the \inhf \bdec{} is 0.24 Bq/g In).
Furthermore, the endpoint energy of \cdhdz is only about 320 keV compared to 496 keV from \inhf. Additionally,
its implementation in a semiconductors implies a better energy resolution than
scintillators. Disadvantages are the smaller natural abundance 
with respect to \inhf, being a factor eight less, and the much higher
threshold for allowed transitions. $^{50}$V as the third 4-fold forbidden
beta-decay emitter is not appropriate for low energy solar neutrino searches because of
its low natural abundance and lack of interesting low lying 1$^+$ states.
A comparison of the three isotopes is shown in Tab.~\ref{tab:fourfold}.

\subsection{Real time pp-detection using of \tehfz}
A chance for real-time pp-neutrino detection exists using two allowed states in \tehfz (Fig~\ref{}). 
The first excited 3/2$^+$ state
in \ihfz is 188 keV above the ground state allowing pp-detection
with a threshold of \enu = 366 keV. The coincidence will be formed by a double tag of an electron
in direct coincidence to the emission of a 188 keV photon or the corresponding cascade. An additional 1/2$^+$ state
exist with a threshold of \enu = 420 keV resulting in the emission of a 243 keV photon. Additionally two further 
3/2$^+$ states can be used for \bes detection having thresholds of 549 keV and 631 keV. 
Associated with them is the emision of 453.8 or 372.1 keV photons respectively.

\subsection{Measuring solar antineutrinos}
The existence of active flavours coming from the Sun  besides \nel
is established by SNO.
However, it is an open question what this new flavour actually is.
Therefore a measurement of a
possible solar antineutrino flux might be useful.
In CdTe (CdZnTe) there are  3 (4) isotopes available of various forms of \bpbp decay, allowing for a tag on possible
solar antineutrinos. 
From a principle point of view no
antineutrino below 511 keV can be detected via charged current reactions
on nuclei because one has to account for the positron mass. Therefore pp-antineutrinos cannot be
observed by this method.
The most promising candidate in CdTe is \cdhs, allowing antineutrino detection with a threshold of
\enu = 713 keV.
The tag would be a monoenergetic 149 keV positron together with a decay of
$^{106}$Ag via $\beta^+$ or electron capture 
and a half life of 24 min.

\section{Experimental considerations}
In the following the detection of the coincidences is discussed close to
the
design presented in
\cite{cobra}. It is assuming an array of CdTe detectors each of 1cm$^3$ size. Such a design has a large
advantage for coincidence searches. The signals containing two electron tags have to rely on the
fact, that the same crystal has to fire twice. The individual rates of such a crystal can be small
and applying corresponding energy cuts appropriate for the tag will reduce
background
significantly. Furthermore in case of monoenergetic electrons as expected from \bes \neus the good
energy resolution allows to define a rather tight constraint on the coincidence signal.
To avoid background from \cdhdz \bdec{} the second electron can be required to have at least 320 keV.
This is completely reducing this background, while keeping the efficiency still high, because the
interesting \bdec s have Q-values well above 2 MeV (e.g. the \bdec{} of
\inhs has
a Q-value of 3.27 MeV). None of the
$\gamma$-lines to be observed from the signal are in close
vicinity of the strongest lines of the natural decay chains. The most dangerous might be an effect
of the 46.5 keV line of $^{214}$Bi on the 39.9 keV IT line occuring in the \tehd tag. Also the 672 keV
line of the \cdhdz tag is in vicinity of the 662 keV line of $^{137}$Cs,
but both can be significantly 
reduced by the coincidence techniques and the good energy resolution.\\
 For higher energetic
gammas the coincidence of neighbouring
crystals has to be used, because no delayed coincidences can be formed. The efficiency of gamma-rays 
leaving the crystal without further interactions is
increasing with energy and is beyond 60 \% already at 250 keV as obtained by a GEANT4 Monte Carlo simulation
\cite{hk}. Another
step forward in signal
identification would be the usage of pixelised detectors, which would
constrain the vertex for two electron tags to one pixel and 
additionally two tracks consistent with electrons have
to start from that pixel. Also multiple interactions within a crystal
like an electron together with a gamma, can be probed in that way. Even without pixel detectors such
a discrimination seems possible by pulse shape analysis.\\
For calibration purposes several solar neutrino experiments have used MCi $^{51}$Cr source, producing
monoenergetic lines of 743 keV (90 \%) and 426 keV (10 \%) neutrinos
\cite{kirst}. Two more sources of $^{75}$Se with \enu = 451/461 keV and
$^{37}$Ar with \enu = 814 keV are also under consideration \cite{kom,gav}. The $^{51}$Cr 
source would be appropriate here as well, because
all nuclear levels discussed as signal levels except one of the excited 1$^+$ states in \tehd and one of the
excited 3/2$^+$ states in \cdhdz can be populated by
the source. The latter depends on the precise Q-value of the \cdhdz \bdec .

\section{Rates}
Observed rates can be determined for the ground state transitions by using the known ft-values
of the corresponding $\beta$-decays. The cross-section can be determined via the relation
\cite{bv}
\begin{equation}
\sigma = \frac{2.64 \cdot 10^{-41}}{ft} \frac{2I'+1}{2I+1} p_e E_e F(Z,E_e) \quad \mbox{(cm$^2$)}
\end{equation}
where $I',I$ are the involved nuclear spins, $p_e$ and $E_e$ are the
momentum and energy of the outgoing electron in units of the
electron mass and F(E,Z) is the Coulomb function. 
The used ft-values of \bdec{} are
taken from \cite{ftv,bat98}. 
The Gamow-Teller transition
matrix elements for \cd were measured recently \cite{akimune}. The expected rates
from \bes only are 89 SNU for \cd and 10 SNU 
for \zns. For the excited states transitions the
GT matrix elements have
to be measured or calculated and without their knowledge, rates cannot be seriously predicted. 
Therefore the above mentioned neutrino sources are very important. 
Accelerator measurements can be done using charge exchange reaction like (p,n) or ($^3$He,t) \cite{ejiri}. 
A more sophisticated analysis wil aslo include efficiencies from Monte Carlo simulations and
details on the $\gamma$-emission in the nuclear de-excitation.

\section{Summary and conclusions}
The prospects of various isotopes of Cd, Zn and Te for low energy solar
neutrino spectroscopy are explored. To obtain a reasonable signal various
coincidence tags can be used, as compiled in Tab.~\ref{tab:comparison}. It allows the detection of
$^7$Be in real
time for five isotopes and therefore offers redundancy in the obtained results. 
The most promising detection signal is the ground state transition 
of \cd to \inhs resulting in 89 SNU. This has to be seen as a lower limit because CNO contributions are
not taken into account.
In addition $^{125}$Te allows a real time detection of pp-neutrinos with a threshold of
330 keV. The usage of semiconductors is
advantageous for background reduction for $^7$Be detection
is  because the monoenergetic electron forming the first step of the coincidence can
be measured with good precision. Rates for excited state transitions cannnot be determined reliably because
a lack of knowledge in the corresponding GT matrix elements, a problem also known from other low energy
solar neutrino experiments. It might be worthwile to consider an experimental program to measure these
matrix elements, which would also be valueable for double beta decay. As common for solar neutrino detection
detector sizes of tons have to be considered, this kind of experiment is 
not feasable in the very near future.

\section{Acknowledgements}
I would like to thank H. Ejiri, Y. Ramachers and S. Schoenert for
valueable discussions and comments.
This work is supported by a Heisenberg-Fellowship of the Deutsche Forschungsgemeinschaft.

\newpage

\clearpage
\newpage

\begin{center}
\begin{table}[hhh]
\begin{tabular}{|c|c|c|c|c|}
\hline
Isotope & nat. ab. (\%) & half-life (yrs) & Q-value (keV) & \enu Thr. (keV)\\
\hline
$^{50}$V & 0.25 & $1.4 \cdot 10^{17}$ & 1038 & 2126\\
$^{113}$Cd & 12.2 & $9 \cdot 10^{15}$ & 320 & 709 \\
$^{115}$In & 95.7 & $4.4 \cdot 10^{14}$ & 496 & 118 \\
\hline
\end{tabular}
\bigskip\\
\caption{\label{tab:fourfold} Comparison of the three known 4-fold forbidden \bdec{} isotopes
and their potential for real time solar neutrino spectroscopy. Shown
are the natural abundance, \bdec{} half-life, Q-value of the \bdec{} and the
threshold for solar neutrino detection.}
\end{table}
\end{center}

\begin{center}
\begin{table}[hhh]
\begin{tabular}{|c|c|c|c|} 
\hline
Isotope & nat. ab. (\%) & \enu Thr. (keV) & solar $\nu$ sources\\ 
\hline
\zns  & 0.62 & 655 & \bes,$^{13}N,^{15}O,^{17}F,^8B$,pep\\
\cdhvz & 28.7 & 1440 & $^{15}O,^{17}F,^8B$,pep\\
\cd & 7.5 & 464 & \bes,$^{13}N,^{15}O,^{17}F,^8B$,pep \\
\teha & 31.7 & 1258 & $^{15}O,^{17}F,^8B$,pep\\
\tehd & 33.8 & 494 & \bes,$^{13}N,^{15}O,^{17}F,^8B$,pep \\
\hline
\end{tabular}
\bigskip\\ 
\caption{\label{tab:comparison} Compilation of double beta isotopes in CdZnTe and their
sensitivity
to various components of the solar neutrino flux. Shown are the natural abundances,
the solar neutrino energy threshold and the contributing \neu fluxes.}
\end{table}
\end{center}

\end{document}